\documentstyle[11pt,epsfig]{article}
\newcommand{\beq}{\begin{equation}}
\newcommand{\eeq}{\end{equation}}
\newcommand{\bea}{\begin{eqnarray}}
\newcommand{\eea}{\end{eqnarray}}
\def\lsim{\raise0.3ex\hbox{$\;<$\kern-0.75em\raise-1.1ex\hbox{$\sim\;$}}}
\def\gsim{\raise0.3ex\hbox{$\;>$\kern-0.75em\raise-1.1ex\hbox{$\sim\;$}}}
 
\begin{document}
\begin{center}
\large{\bf Galactic PeV Neutrinos} \\

\medskip
{Nayantara Gupta
 \footnote{nayan@rri.res.in}\\
Raman Research Institute, Sadashiva Nagar, Bangalore 560080, India }
\end{center}
\begin{abstract}
The IceCube experiment has detected two neutrinos with energies beween 1-10 PeV.
They might have originated from Galactic or extragalactic sources of cosmic rays. In the present work we consider hadronic interactions of the diffuse very high energy cosmic rays with the interstellar matter within our Galaxy to explain the PeV neutrino events detected in IceCube. We also expect PeV gamma ray events along with the PeV neutrino events if the observed PeV neutrinos were produced within our Galaxy in hadronic interactions. PeV gamma rays are unlikely to reach us from sources outside our Galaxy due to pair production with cosmic background radiation fields. We suggest that in future with simultaneous detections of PeV gamma rays and neutrinos it would be possible to distinguish between Galactic and extragalactic origins of very high energy neutrinos.   
\end{abstract}
\section{Introduction}
 The field of neutrino astronomy has been enriched with many theoretical predictions of astrophysical neutrinos in the past few decades and the neutrino physicists are eagerly waiting for the detection of these neutrinos. The two neutrino events with energies between 1 to 10 PeV detected in IceCube on August 9, 2011 and January 3, 2012 could be of atmospheric or astrophysical origin \cite{aartsen}. If the two PeV neutrino events detected in IceCube are of astrophysical origin then more events are expected to be detected in the near future.  
\par
The events could have originated from GRBs as discussed in \cite{cholis}, or low luminosity GRBs and Pop III GRBs \cite{liu}. Cosmogenic neutrinos produced in interactions of ultrahigh energy cosmic rays with cosmic background radiations are expected to be more abundant at higher energies. The Glashow resonance events from cosmogenic neutrinos may give showers in IceCube detector \cite{barger}. However, in another study \cite{roulet} the authors have argued that the two events detected by IceCube are hard to explain with cosmogenic neutrino flux.   
The detection of electron type shower events and non detection of muon tracks in IceCube may be explained with cosmogenic neutrinos including the effects of Glashow resonance and neutrino decay or Lorentz violation as suggested in \cite{arti}. 
\par
The IceCube detected TeV and PeV energy events have been analysed in some more recent papers \cite{laha,anch3,he} and the authors have discussed various possible origins of these events.

\par
It has been discussed  earlier \cite{gupta1} that the secondary gamma ray and neutrino flux produced in interactions of very high energy cosmic rays (VHECRs) ($\geq 10^{16}$ eV) propagating within our Galaxy could be useful to study the highest energy cosmic accelerators. 
In this paper we have shown that the two PeV neutrino events detected in IceCube could have their origin within our Galaxy in interactions of VHECRs with the interstellar matter. A detailed study combining all the TeV-PeV events will be presented in near future \cite{joshi}.  
\section{Neutrino Events from Hadronic Interactions of VHECRs}
Very high energy cosmic rays interact with matter and radiation fields during their propagation. Cosmic ray protons produce secondary pions interacting with the cold matter $p-p$ and background photons $p-\gamma$ in our Galaxy. The time scale of $p-p$ interactions is much smaller compared to $p-\gamma$ interactions in the infrared background of our Galaxy. This has been shown in \cite{gupta1} with the infrared photon density of our Galaxy from \cite{mos}. Heavy nuclei also produce secondary gamma rays in pure hadronic $Fe-p$ interactions and photo-disintegration followed by de-excitation of daughter nuclei. The time scales of these processes have been compared in Fig.1. of \cite{gupta1}. In the energy range of our interest the pure hadronic interaction $Fe-p$ has a shorter time scale than photo-disintegration of heavy nuclei.
The very high energy photon flux produced in $pp$ interactions and subsequent $\pi^0$ decay is
\beq
F_{\gamma,pp}(E_{\gamma})= \frac{t_{esc,p}}{t_{pp}}G_{pp}(E_{\gamma})Y_{\alpha},
\label {gamma_p}
\eeq
where $E_{\gamma}=0.1 E_p$. $t_{esc,p}$ and $t_{pp}=1/(c\sigma_{pp}n_H)$ are the escape and $pp$ interaction time scales for $\pi^0$ production of cosmic ray protons respectively, $n_H$ is number density of ambient hydrogen molecules in the Milky Way. The cross section for $pp$ interactions is included in $t_{pp}$.
 The proton flux has been included within $G_{pp}(E_{\gamma})$.
\beq
G_{pp}(E_{\gamma})=2\int_{E_{\pi^0,min}}^{E_{\pi^0,max}}\frac{dN_p(E_{\pi^0})}{dE_{\pi^0}}\frac{dE_{\pi^0}}{(E_{\pi^0}^2-m_{\pi^0}^2)^{1/2}},
\label{g_pp}
\eeq
where $\frac{dN_p(E_{\pi^0})}{dE_{\pi^0}}=A_p E_{\pi^0}^{-\alpha}$, $A_p$ is normalisation constant and $\alpha$ is spectral index of the proton spectrum, $E_{\pi^0,min}=E_{\gamma}+m_{\pi^0}^2/(4E_{\gamma})$. The maximum energy of pions is the maximum energy of the cosmic ray proton/nucleon $E_{\pi^0,max}=E_n^{max}$.
  In $pp$ interactions $\pi^0$ carries on the average $20\%$ of the cosmic ray proton's energy. The spectrum-weighted moments $Y_{\alpha}$ has been calculated 
below 
\beq
Y_{\alpha}=\int_0^1 x^{\alpha-1}f_{\pi^0}(x) dx.
\label{y_l}
\eeq
The function $f_{\pi^0}(x)\simeq8.18x^{1/2}\Big(\frac{1-x^{1/2}}{1+1.33x^{1/2}(1-x^{1/2})})\Big)^4\Big(\frac{1}{1-x^{1/2}}+\frac{1.33(1-2x^{1/2})}{1+1.33x^{1/2}(1-x^{1/2})}\Big)$ with $x=E_{\pi^0}/E_p$.  
The gamma ray flux $F_{\gamma,Ap}(E_{\gamma})$ produced in pure hadronic interactions of cosmic ray nuclei is 
\beq
F_{\gamma,Ap}(E_{\gamma})=\frac{t_{esc,A}}{t_{hadr}}G_{Ap}(E_{\gamma})Y_{\alpha},
\label{gamma_fe}
\eeq
where $t_{esc,A}$ is the escape time scale of cosmic ray nuclei of mass number $A$. The flux of iron nuclei can be expressed as a power law in energy similar to the proton flux $\frac{dN_{FE}(E_{FE})}{dE_{FE}}=A_{FE}E_{FE}^{-\alpha}$, where $E_{FE}=56 E_p$. The expression for $G_{Ap}$ for iron nuclei can be calculated similar to $G_{pp}$, except the proton flux is to be replaced by flux of iron nuclei per unit nucleon energy. 

\par
  The hadronic interaction time scale of cosmic ray nuclei can be estimated with hadronic interaction cross section for $\pi^0$ production 
$\sigma_{hadr}=\sigma_{pp}A^{3/4}mb$ \cite{leb}, where $A$ is mass number of nuclei. The energy dependent expression for $pp$ interaction cross section is 
$\sigma_{pp}=34.3+1.88 \ln(E_p/1TeV)+0.25 (\ln(E_p/1TeV))^2 mb$ \cite{kelner}.
\par
We have also checked with the distribution function given in \cite{kelner} for calculation of gamma ray flux using their eqn.(72). We find the fluxes calculated using the spectrum weighted momenta in eqn.(\ref{y_l}) and with the distribution function given in \cite{kelner} are comparable. 
\par
The escape time scales of VHECR protons and nuclei are not known. In principle one may vary the escape time scale to fit the observed gamma ray flux. 
The ratio of the fluxes of gamma rays and cosmic ray nuclei of mass number $A$ at the {\bf same energy} can be expressed as
\beq
\frac{F_{\gamma,Ap}}{F_{CR,A}}=\frac{t_{esc,A}}{t_{hadr}}\frac{2 Y_{\alpha}}{\alpha} A^{-\alpha+1}
\label{r_fluxes}
\eeq
The ratio of the fluxes is same as above as the gamma ray and cosmic ray fluxes have the same energy dependence $E^{-\alpha}$.
The contribution to the secondary gamma ray and neutrino flux from heavier nuclei cosmic rays is expected to be lower than that from cosmic ray protons.
\par
The neutrino fluxes of each flavor (after adding neutrino and antineutrino fluxes) expected on earth are similar to the gamma ray flux as discussed in \cite{gaisser,cost,kappes,anch2,kist}. Thus at any energy $\nu_e:\nu_{\mu}:\nu_{\tau}:\gamma \sim 1:1:1:1$. We note that pion and muon cooling are not important in Galactic magnetic field which affect the neutrino flavor ratios \cite{kashti} to be detected on earth from astrophysical sources.
Eqn.(6) can also be used to find the ratio of fluxes of neutrinos and cosmic rays as neutrinos of each flavor and gamma rays are produced equally. 
\par 
CASA-MIA experiment \cite{casa} set upper limits on the ratio of isotropic diffuse gamma ray and cosmic ray fluxes in the energy range of 0.57 PeV to 55 PeV.
In their Table 2 they have listed the mean gamma ray and cosmic ray energies and the ratios of their integral fluxes. The mean gamma ray and cosmic ray energies differ by less than a factor of 2. Within the large energy range of 0.57 PeV to 55 PeV the upper limit on the ratios of gamma ray and cosmic ray integral fluxes remains approximately constant at $10^{-4}$.  
\par
Recently, IceCube has presented their preliminary result on the ratio of gamma ray and cosmic ray fluxes from the Galactic Plane region \cite{ice_pev}. The upper limit on the ratio in the energy range of 1.2-6 PeV has been found to be $1.2\times 10^{-3}$ with $90\%$ confidence level in their work. After 5 years of operation the completed detector is expected to reach more than an order of magnitude better sensitivity. The upper limit obtained by CASA-MIA \cite{casa} is more conservative than the current limit of $1.2\times 10^{-3}$ set by IceCube.   
With the CASA-MIA upper limit and the observed cosmic ray flux at 1 PeV energy we obtain the expected flux of neutrinos at 1 PeV. 
Please note that we are not taking the ratio of the neutrino flux to their parent cosmic ray flux. We are interested in the ratio of the gamma ray or neutrino flux to the cosmic ray flux at the same or comparable energy to compare it with the CASA-MIA upper limit. 
\par
The observed flux of VHECRs near 1 PeV is $\sim 10^{-4} GeV/cm^2/sec/sr$.
If we apply the CASA-MIA upper limit $10^{-4}$ then we expect neutrino flux of each flavor $10^{-8} GeV/cm^2/sec/sr$ at 1 PeV. 
The total neutrino flux after adding up the contributions from the three flavors is expected to be $3\times10^{-8} GeV/cm^2/sec/sr$ which is comparable to the
upper limit on the total neutrino flux $3.6\times10^{-8} GeV/cm^2/sec/sr$ obtained by IceCube collaboration \cite{abbasi}.
We note that from our eqn.(6) one can estimate the escape time scale of PeV cosmic rays using the CASA-MIA upper limit of $10^{-4}$. If we assume the PeV cosmic rays are protons then we get from eqn.(6) their escape time scale is $10^{14}sec$ assuming number density of ambient hydrogen molecules $n_H=1$ and differential spectral index of cosmic ray spectrum $\alpha=3$.
\par
The interaction lengths of the various neutrino interactions (charged current, neutral current and Glashow resonance) are given in \cite{rg}.
The charged current cross-section can be expressed as a function of neutrino energy \cite{rg} $\sigma_{CC}\simeq 5.53\times 10^{-36} \Big(E_{\nu}/GeV\Big)^{0.363}cm^2$. 1-10 PeV energy range is the ideal energy window to study astrophysical neutrinos as the Glashow resonance effect $\bar\nu_e+ e^-\rightarrow hadrons$ at 6.3 PeV enhances the neutrino signals. We note the cross-section of this channel \cite{glashow} is $3.41 \times 10^{-31} cm^2$ from Table 3 of \cite{raj}. Due to the smaller interaction length of the Glashow resonance interactions PeV electron antineutrino showers are much more likely to be detected in IceCube compared to neutrino events from charged and neutral current interactions. However, the analysis presented by IceCube Collaboration \cite{aartsen} shows the detected events are unlikely to be of Glashow resonance events. They might have originated from charged current interactions of electron flavor neutrinos or antineutrinos. 
\par
Within the scenario discussed in the present work PeV gamma rays are also expected to be produced with neutrinos. The PeV gamma rays might be detected in future with higher IceCube sensitivity. Extragalactic PeV gamma rays are unlikely to reach us due to pair production losses in cosmic background radiation fields. 
The mean free pathlength for pair production ($e^-e^+$) for PeV gamma rays with cosmic microwave background photons is 10kpc \cite{pb}.
The simultaneous search for PeV gamma rays and neutrinos with IceCube would provide unique opportunity to understand whether the PeV neutrinos have Galactic or extragalactic origin. 

\section{ Conclusions}
We have shown that the interactions of VHECRs with Galactic matter may explain the PeV neutrino events observed in IceCube. We propose that in future simultaneous searches for PeV gamma rays and neutrinos would be useful to distinguish between Galactic and extragalactic sources of neutrinos.   

\end{document}